# Neutron diffraction study of YVO$_3$, NdVO$_3$, and TbVO$_3$

M. Reehuis, [1,2] C. Ulrich, [1] P. Pattison, [3], B. Ouladdiaf, [4] M.C. Rheinstädter, [4] M. Ohl, [5] L.P. Regnault, [6] M. Miyasaka, [7] Y. Tokura, [7,8] and B. Keimer [1]

[1]*Max-Planck-Institut für Festkörperforschung, D-70569 Stuttgart, Germany*

[2]*Hahn-Meitner-Institut, D-14109 Berlin, Germany*

[3] *SNBL at ESRF, BP 220, F-38042 Grenoble Cedex 9, France*

[4]*Institut Laue-Langevin, BP 156, F-38042 Grenoble Cedex 9, France*

[5] *Forschungszentrum Jülich GmbH, D-52425 Jülich, Germany*

[6] *CEA Grenoble, Département de Recherche Fondamentale sur la Matière Condensée, 38054 Grenoble cedex 9, France*

[7]*Department of Applied Physics, University of Tokyo, 113 Tokyo, Japan*

[8] *Correlated Electron Research Center (CERC), National Institute of Advanced Industrial Science and Technology (AIST), Tsukuba 305-8562, Japan*

The structural and magnetic properties of YVO$_3$, NdVO$_3$ and TbVO$_3$ were investigated by single-crystal and powder neutron diffraction. YVO$_3$ shows a structural phase transition at 200 K from an orthorhombic structure with the space group *Pbnm* to a monoclinc phase. The diffraction pattern in the monoclinic phase is well described by the space group $P2_1/b$, but supplementary high-resolution x-ray diffraction experiments showed that the monoclinic distortion is extremely small. The magnetic order of the vanadium sublattice with an ordered moment much lower than the free-ion value sets in at 116 K. In the monoclinic phase the magnetic ordering pattern can be described with the basis function [$C_x$, $C_y$, $G_z$]. A group theoretical analysis shows that this magnetic state is incompatible with the lattice structure, unless terms of higher than bilinear order in the spin operators are incorporated in the spin Hamiltonian. This observation is discussed in the light of recent theories invoking unusual many-body correlations between the vanadium $t_{2g}$ orbitals. A structural phase transition back to the orthorhombic space group *Pbnm* is observed upon cooling below 77 K. This transition is accompanied by a rearrangement of the magnetic structure into a pure $G_z$ mode with an ordered moment close to the free-ion value. Refinements of the diffraction patterns revealed the atomic positions and bond distances in all three structural phases. The crystal structures of NdVO$_3$ and TbVO$_3$ are closely similar to that of YVO$_3$. However, only a single magnetic phase transition into a phase characterized by the modes $C_x$ and $C_y$ was found in the vanadium sublattice down to 9.5 K. Below 60 K the magnetic moments of the Nd$^{3+}$- and Tb$^{3+}$-ions are gradually polarized by the antiferromagnetically ordered vanadium moments. Below 11 K, we found a non-collinear order of the terbium moments with the modes $C_x$ and $F_y$.

PACS number(s): 61.10.Nz, 61.12.Ld, 61.50.Ks, 75.25.+z, 75.50.Ee

## I. INTRODUCTION

The perovskite structure class with $ABX_3$ stoichiometry ($A$ = larger cation, $B$ = smaller cation, $X$ = anion) has attracted considerable attention in recent years. Because of the great variety of possible elements, a huge number of compounds of this type are known. The majority of them are transition metal oxides and fluorides. Some members of this familiy show interesting physical properties such as the Mott transition, colossal magnetoresistance, ferroelectricity, as well as charge, orbital and spin ordering. Orthovanadates of chemical composition $R$VO$_3$, with $R$ = Y or trivalent rare-earth metal, are particularly interesting examples, as they undergo multiple orbital and magnetic ordering transitions as a function of temperature. Further, insulator-metal transitions are observed in these systems upon substitution of $R$ by divalent cations such as $Sr^{2+}$.

The ideal perovskite structure belongs to the cubic space group $Pm-3m$, where the $B$-cations are octahedrally coordinated by six $X$-anions building a three-dimensional network of corner-sharing $BX_6$-octahedra. However, most perovskites have tilted and/or distorted $BX_6$-octahedra leading to a lower symmetry. [1–4] At room temperature, the principal cause of the deviations from the ideal perovskite structure is the influence of the relative ionic sizes. This "ionic size effect" leads to tilts of the octahedra, as shown in Fig. 1 for YVO$_3$. The room temperature crystal structure of the vanadates $R$VO$_3$ is thus described by the orthorhombic space group $Pbnm$. In Fig. 2, the unit cell parameters taken from Ref. 5 are plotted against the atomic number of the $R$-atoms. From La to Er the cell volume decreases continuously due to the well-known lanthanide contraction. However, only the lattice constants $a$ and $c$ decrease significantly, whereas $b$ increases slightly. This is due to the fact that in the case of the smaller $R^{3+}$-ions the VO$_6$-octahedra are more strongly rotated around the $b$- and $c$-axes than along the $a$-axis of the orthorhombic structure. The materials with the largest $R^{3+}$-ions, LaVO$_3$ and CeVO$_3$, exhibit pseudocubic room-temperature lattice structures with $a \sim b \sim c / \sqrt{2}$.

The nearly cubic crystal field splits the degenerate manifold of atomic $d$-electrons on the $B$-cations into a lower-lying triplet of $t_{2g}$ symmetry and a high-energy doublet of $e_g$ symmetry, with an energy difference that typically exceeds 1 eV. Further distortions away from cubic symmetry split the residual degeneracy of the sub-manifolds. For cations with incompletely filled $d$-electron shells, electronic energy can thus be gained by a cooperative distortion of the octahedra through the *Jahn-Teller* (*JT*) effect, which sets in below an "orbital ordering" temperature. The relative orientation of orbitals on neighboring cations then determines the superexchange interactions between the unpaired *d*-electrons, which in turn determine the magnetic ordering patterns. For perovskites with partially occupied $e_g$ orbitals, such as cuprates and manganites, the *JT* energy greatly exceeds the superexchange coupling between unpaired electrons, so that orbital and magnetic ordering temperatures are well separated. The two valence electrons of the $V^{3+}$ ions in the vanadates, however, occupy the $t_{2g}$ orbitals (*xy*, *yz*, and *xz*). Owing to the higher degeneracy and weaker *JT* coupling of the $t_{2g}$ states, these systems undergo orbital and magnetic ordering transitions in the same temperature range. Magnetic and orbital degrees of freedom are thus inextricably intertwined.

The magnetic and orbital ground states of $YVO_3$, in particular, have been the subject of much attention. Investigations of the structural and magnetic properties of this compound started in the 1950's.[6–8] Antiferromagnetic ordering of the vanadium moments was first observed below about 110 K.[9] In later studies, two magnetic phase transitions were identified at $T_1$ = 78 K and $T_2$ = 114 K.[10-12] Specific heat measurements showed that the transition at $T_1$ is of first and that at $T_2$ of second order. Further, neutron powder diffraction showed a change of the magnetic order of the $V^{3+}$-sublattice from a *C*-type to a *G*-type with decreasing temperature at $T_1$, whereas another paper[13] claimed a transition from a *G*-type to a *C*-type. Kawano *et al.*[14] finally confirmed a *C*-type ordering below $T_2$ and a transition into a *G*-type below $T_1$. Recently we reinvestigated the magnetic structure of $YVO_3$ by single-crystal neutron diffraction.[15] Below 77 K a pure $G_z$-type was confirmed, while in the intermediate phase a noncollinear structure was discovered. Here, a weak *z*-component of the *G*-type was observed along with the previously reported strong *C*-modes in the *ab*-plane. In addition to the change of magnetic ground state, the transition at $T_1$ is associated with a modification of the lattice structure, and with profound changes in the magnetic excitation spectrum and the optical properties. Based on these observations, it has been identified as an orbital ordering transition. Another structural transition at $T_3$ = 200 K also leads to a modification of the optical spectrum and has also been attributed to orbital ordering. In the magnetically ordered phases, multiple magnetization reversals and unusual magnetization memory effects are observed.

The nature of the orbitally ordered phases of $YVO_3$, the microscopic mechanism of the orbital ordering transitions, and the origin of the magnetization memory effects are still poorly understood. It is generally believed that the degeneracy of the $t_{2g}$ levels is successively lowered upon cooling below the structural phase transitions at $T_3$ and $T_1$, and that the degeneracy is completely lifted at low temperatures. The orbital ordering pattern below $T_1$ is also relatively uncontroversial. In the low-temperature phase, the vanadium *xy* orbital is believed to be occupied by one of the two valence electrons, while the other electron occupies the *xz* and *yz* orbitals in an alternating fashion in the *ab*-plane. Along the *c*-axis direction, the occupied *xz* and *yz* orbitals are uniformly stacked (*C*-type orbital order). The classical Goodenough-Kanamori rules yield a consistent description of this orbital state as well as the magnetic ground state and excitations. In contrast, the microscopic magnetic properties of the intermediate-temperature phase ($T_1 < T < T_3$) are difficult to describe in this classical picture, and several different proposals have been made for the orbital and magnetic ground state in this temperature range. [16-21] An intriguing aspect of some of these proposals is the existence of strong quantum fluctuations in both spin and orbital sectors. While these fluctuations may be responsible for the anomalously small ordered moment and other unusual features of the intermediate phase, direct evidence has thus far proven elusive.

Miyasaka et al. recently studied the evolution of the sequence of phase transitions with the size of the $R^{3+}$-ion. [22] While the lowest-temperature transition at $T_1$ is only observed for the smallest cations, the magnetic transition at $T_2$ and structural transition at $T_3$ evolve smoothly as a function of ionic size. For the largest cations, $Ce^{3+}$ and $La^{3+}$, the sequence of magnetic and structural transitions was reported to be reversed, which implies a magnetically ordered "orbital liquid" phase in a narrow temperature range above the Néel transition. The low-temperature phase of the end member $LaVO_3$ exhibits *C*-type magnetic order [23,24], similar to the intermediate-temperature phase of $YVO_3$.

The purpose of this paper is to provide a more complete basis for the theoretical description of the electronic properties of the vanadates. To this end, we have used single-crystal neutron diffraction to carry out a comprehensive investigation of the crystal structures and magnetic ground states of $YVO_3$ as well as its sister compounds $TbVO_3$ and $NdVO_3$, which undergo different sequences of phase transitions as a function of temperature. In the latter compounds, the exchange interactions between the Tb- and Nd-moments, respectively, and the V-moments were also studied. The paper is organized as follows. Section II contains a description of the experimental setup. In Section III, we present data on the crystal structures (part A) and magnetic structures (part B) of the title compounds and discuss these in the light of prior studies in and the literature. Finally, Section IV provides a brief summary of the major conclusions.

## II. EXPERIMENTAL DETAILS

For our experiments cylindrical crystals of $YVO_3$, $NdVO_3$ and $TbVO_3$ with dimensions $d$ = 4 mm and $h$ = 5 mm were used. The crystals were grown by a floating-zone method as described elsewhere.[25] All crystals were untwinned. Single-crystal neutron diffraction data were collected on the four-circle diffractometer E5 at the BER II reactor of the Hahn-Meitner-Institut in Berlin. The crystal structures were determined from data sets using the neutron wavelength $\lambda$ = 0.89 Å (Cu-monochromator). The refinements of the crystal structures were carried out with the program *Xtal* 3.4.[26] Here the nuclear scattering lengths $b$(O) 5.805 fm, $b$(V) = –0.3824 fm, $b$(Y) = 7.75 fm, $b$(Nd) = 7.69 fm, and $b$(Tb) = 7.38 fm were used.[27] For the investigation of the magnetic structure, we collected data sets using a pyrolytic-graphite monochromator selecting the longer neutron wavelength $\lambda$ = 2.4 Å. At 150 K, well above the Néel temperature, we collected data sets of $YVO_3$, $NdVO_3$ and $TbVO_3$ in order to determine the overall scale factor from the crystal structure refinements. With the absorption and extinction corrected magnetic structure factors, we were able to obtain the magnetic moments of the metal ions in the magnetically ordered range. The moments of the V-, Nd- and Tb-atoms were refined with the program *FullProf*.[28] The magnetic form factors of the $V^{3+}$-, $Nd^{3+}$, and $Tb^{3+}$-ions were taken from Ref. 29.

Additional measurements of the crystal and magnetic structures of $YVO_3$ were carried out by neutron powder diffraction on the instruments D1A and IN22 at the Institut Laue-Langevin in Grenoble, France. Powder patterns were collected on D1A at different temperatures in the $2\theta$ range between 5 and 150 °. This instrument uses germanium single crystals as monochromator, with the (1 1 5) reflection selecting a neutron wavelength $\lambda$ = 1.9114 Å. Single crystal data on IN22 were taken with pyrolytic-graphite monochromator and analyzer selecting a neutron wavelength of $\lambda$ = 2.4 Å, no collimators, and two pyrolytic-graphite filters. In order to come to a conclusive answer about the crystal symmetry of $YVO_3$, we have performed high-resolution powder diffraction studies using synchrotron radiation at the European Synchrotron Radiation Facility in Grenoble, France. Complete powder patterns on the BM1B (Swiss-Norwegian) beam line with a $2\theta$ range between 1 and 60 ° were collected. The wavelengths used were $\lambda$ = 0.68957 Å at 25 K and 150 K, $\lambda$ = 0.79948 Å at 85 K, and $\lambda$ = 0.49950 Å at 300 K. Rietveld refinements of the powder diffraction data were carried out with the program *FullProf* using the atomic scattering factors provided by the program for the *x*-ray data and the neutron scattering lengths given above.

## III. RESULTS AND DISCUSSION

### A. Crystal structure of YVO$_3$ and TbVO$_3$

We have carried out a comprehensive investigation of the crystal structure of YVO$_3$ at 5, 85, 230, and 295 K. For comparison, the crystal structure of TbVO3 was also investigated at 295 K. We first describe the results of single-crystal neutron diffraction experiments on both compounds at room temperature. A total of 565 and 429 averaged independent structure factors were collected for the yttrium and terbium vanadate, respectively. Refinements of the diffraction patterns were performed in the orthorhombic space group *Pbnm* (standard setting *Pnma*), using the atomic parameters of GdFeO$_3$ as starting values. [30] The positional parameters of the Y- or Tb-, O1- (Wyckoff position 4*c*: *x*, *y*, ¼) and O2-atoms (Wyckoff position 8*d*: *x*, *y*, z) as well as the anisotropic thermal parameters could be refined successfully. Due to the small scattering power of the vanadium atoms [in the position 4*b* (½, 0, 0)] an isotropic thermal parameter $U_{is}$ = 0.003 Å$^2$ was fixed, and it was not allowed to vary during the refinements. The refinements of a total of 23 parameters of YVO$_3$ and TbVO$_3$ resulted in residuals $R(F)$ = 0.028 ($wR$ = 0.031) and $R(F)$ = 0.031 ($wR$ = 0.032). The refined atomic parameters of the yttrium compound are in good agreement with the values obtained recently by Blake et al. [31]

In Table I the parameters of YVO$_3$ are compared with those of TbVO$_3$. Here it can be seen that the positional parameters of the O1- and O2-atoms in the yttrium compound deviate more strongly from the ideal sites (0, ½, ¼) and (½, ½, 0) of the undistorted cubic structure. The observation that a smaller $R^{3+}$-ion favours stronger tiltings of the VO$_6$-octahedra indicates the influence of the ionic-size effect. Table II lists the V-O bond distances of YVO$_3$ and TbVO$_3$ as well as the Fe-O distances in PrFeO$_3$ and LuFeO$_3$ as given earlier in Ref. 30. As Fe$^{3+}$ (electron configuration 3$d^5$, high spin) is orbitally nondegenerate, the distortions of the FeO$_6$ octahedra in the ferrates are driven entirely by the ionic-size effect. Since the room-temperature distortions of the VO$_6$ octahedra in the vanadates are comparable to those of the FeO$_6$ octahedra in the ferrates, we conclude that the *JT* effect is inactive at room temperature.

From the room-temperature refinements, we also obtained the occupancies of the atomic positions. Here it could be seen that all atomic positions of both compounds are fully occupied. Under the assumption of full occupancy on the vanadium site, we obtained for YVO$_3$: *occ*(Y) = 0.998(2), *occ*(O1) = 1.001(2), *occ*(O2) = 0.998(2), and for TbVO$_3$: *occ*(Tb) =

1.003(2), $occ$(O1) = 0.999(3), $occ$(O2) = 0.997(2). This highlights the high quality of the single-crystal samples.

As reported earlier, a monoclinic structure of YVO$_3$ is stable between $T_1$ = 77 K and $T_3$ = 200 K.[31] The temperature dependence of the intensities of some prominent reflections shown in Fig. 3 indeed indicates structural changes with an onset at $T_3$. For instance, the reflections 022 and 002 show a strong change of the intensity at 200 K, while no change can be observed for 020 and 202. This indicates a spontaneous shift of the atoms in particular directions. In agreement with Blake et al.[31], we also observed additional reflections for $T_1 < T < T_3$ which are forbidden in the orthorhombic space group *Pbnm*. These measurements confirm the presence of a structural phase transition from an orthorhombic to a monoclinic structure.[31,32] Assuming a the same monoclinic structure with the space group $P2_1/b$ ($P2_1/b$ 1 1) as found in Ref. 33 for LaVO$_3$, the Rietveld refinements resulted at 85 K in a monoclinic angle $\alpha$ = 89.980(3) °. This is in very good agreement with the value $\alpha$ = 89.977(3) ° (at 80 K) given in Ref. 31. In order to check the presence of monoclinic splittings, we recorded high-resolution powder diagrams using synchrotron radiation, but we could not find any direct evidence of splitting of reflections down to 25 K.

Due to twinning problems below the structural phase transition at 200 K, we refined the atomic positions in YVO$_3$ by neutron powder diffraction. Powder patterns were collected at 5 K (Fig. 4) in the low-temperature orthorhombic phase and at 85 K in the monoclinic phase. A third one was collected at 230 K in the high-temperature orthorhombic phase. The additional reflections characteristic of the monoclinic structure were found to be rather weak and could be observed only in the single-crystal diffraction experiments. Thus we first assumed the crystal structure at 85 K to be pseudo-orthorhombic, and we refined it in the space group *Pbnm*. From the refinements we obtained almost the same residuals and similar V-O distances as observed in the high-temperature orthorhombic phase. The V-O22 (and V-O24) distances in the orthorhombic phase at 230 K and 295 K (Table II) are found to be the largest, followed by V-O21 (and V-O23), both almost lying in the *ab*-plane. The apical bond lengths of V-O11 and V-O12 are clearly the smallest.

We then repeated the refinements in the monoclinic space group $P2_1/b$ and found V-O-bond lengths in good agreement with those of Blake et al.[31]. Here the length of V-O22 (and V-O24) is strongly elongated, while it is strongly shortened for V-O21 (and V-O23) reaching almost the same bond lengths of the apical O1-atoms. However, the determination of the positional parameters of the O2 and O3 atoms, which show the strongest shifts at the orthorhombic-monoclinic transition, is problematic within $P2_1/b$, as these parameters are highly correlated, with correlation coefficients between 84 and 97%. (The correlation coefficient between two parameters $p_i$ and $p_j$ is obtained from the covariance, i.e. the second moment with respect to $p_i$ and $p_j$ by normalization corr($p_i$, $p_j$) = cov($p_i$, $p_j$)/[$\sigma(p_i)$ $\sigma(p_i)$], where $\sigma$ denotes the standard deviation.) As a consequence, the standard deviations of the positional parameters of the O-atoms given in Table I as well as the V-O-bond lengths in Table II are much larger than in the orthorhombic representation, where the corresponding correlation coefficients are smaller than 17 %. Since the residuals and $\chi^2$-values are almost identical within the two space groups [$R_F$(orth) = 2.55, $\chi^2$(orth) = 8.46 and $R_F$(mon) = 2.58, $\chi^2$(mon) = 8.36], we regard the determination of the V-O distances within *Pbnm* as more reliable. This would indicate a weak *JT* effect in the monoclinic phase. However, the issue cannot be resolved definitively, because the intensities of the monoclinic reflections *hkl* and *hk–l* cannot be determined separately from powder data due to the very weak deformation of the Bravais lattice. Further, the additional reflections characteristic of the lower symmetry could not be detected in the powder diagrams.

In the monoclinic phase, the vanadium atoms are located in the Wyckoff positions 2*c* (½, 0, 0 and 0, ½, 0) and 2*d* (½, 0, ½ and 0, ½, ½) of the space group *P*2$_1$/*b*. Despite the lower symmetry, the coordinates of these atoms are still the same as in the orthorhombic space group *Pbnm*. Recently, an even lower symmetric structure with the non-centrosymmetric space group *Pb* (*P b* 1 1) was deduced from ellipsometric measurements.[32] The vanadium atoms as well as the yttrium and oxygen atoms are located in the only possible Wyckoff position 2*a* (*x*, *y*, *z* and –*x*, ½ + *y*, *z*). For both atoms V1 and V2 the parameters *x* and *y* are expected to be close to the values ¼ and ¾, respectively; the parameter *z* of V1 is close to zero, while it is close to ½ for V2. In this symmetry, the vanadium atoms form periodic chains along the *x*- and *y*- directions with constant interatomic distances within the chains. This is not the case for the vanadium chains along *z*, where one expects alternating slightly shortened and elongated V1-V2-distances. In fact, the crystal structure in *Pb* allows the possibility of dimerization along *z*, in agreement with the results of neutron spectroscopy measurements that can be consistently interpreted in terms of an "orbital Peierls state".[15] In our experiment, however, the positional parameters of the vanadium atoms could not be refined with sufficient accuracy to resolve this issue, because the neutron scattering power of vanadium is rather weak and the expected distortion is extremely small. The unusual spin correlations observed in the intermediate-temperature phase of YVO$_3$ are therefore at most weakly reflected in the lattice structure. This may be due to the weak lattice coupling of the $t_{2g}$ valence orbitals, in conjunction with strong orbital fluctuations.

The thermal variation of the intensities of the reflections in Fig. 3 clearly indicates a strong change in the crystal structure of YVO$_3$ at about 77 K. At this temperature, Kawano et al. also observed a spontaneous change of the lattice constants. [14] They found that the values *a* and *c* spontaneously increase by about 0.15 - 0.16 %, while *b* decreases strongly by about 0.64 %. Below this transition temperature the crystal structure again can be described in the orthorhombic space group *Pbnm*. [31,32] The results of the Rietveld refinements of the data set collected at 5 K are summarized in Tables I and II. Here it can be seen that the V-O1-bond distances at 5 K do not strongly differ from the values of the monoclinic structure determined at 85 K. But strong shifts could be observed for the O2-atoms. The length of V-O22 (and V-O24), the largest at higher temperature, strongly decreases reaching almost the same value of that of V-O11 (and V-O12). On the other hand the value of V-O21 (and V-O23) is now clearly the largest. This is in contrast to the result of Blake et al. [31] In contrast to the intermediate-temperature phase, the neutron data in the low-temperature phase therefore exhibit an unequivocal signature of the Jahn-Teller effect, in line with theoretical expectations.

## B. Magnetic order in *R*VO$_3$ with *R* = Y, Nd, Tb

We begin this section by presenting magnetic neutron scattering data on YVO$_3$, following our recent brief account of the magnetic properties of this material [15]. In the monoclinic phase between $T_1$ = 77 K and the Néel temperature $T_2$ = 116 K, the strongest magnetic intensities could be observed on the reflections 100 and 010. The relatively strong magnetic intensities of the peaks 102 and 012 indicate that the magnetic moments order predominantly within the *ab*-plane. The magnetic structure in this plane is of the *C*-type, where the spin sequence is + + − − for the V-atoms in the positions ½ 0 0, ½ 0 ½, 0 ½ 0 and 0 ½ ½. At 85 K, the components along the *a*- and *b*-axes are 0.49(3) µ$_B$ and 0.89(2) µ$_B$, respectively. For comparison, the values quoted earlier by Blake et al. at 80 K are 0.649(15) µ$_B$ and 0.895(13) µ$_B$, respectively. [31] In Fig. 4 it can be seen that reflection 101 increases spontaneously with increasing temperature at $T_1$ = 77 K. This suggests an additional magnetic order of the V-moments along the *c*-axis in the *G*-type with the spin sequence + − − +. A corresponding moment of 0.30(4) µ$_B$ could be refined. In the low-temperature orthorhombic phase at 10 K, a pure *G*-type ordering along the *c*-axis with a moment value of 1.72(5) µ$_B$ was found. In contrast to the intermediate-temperature phase, where the total magnetic moment is far below the free-ion value of 2 µ$_B$, the moment in low-temperature phase does not deviate substantially from this value. The residual discrepancy can be attributed to zero-point fluctuations and a small unquenched orbital moment.

In Ref. 15, we had discussed the large noncollinearity and the small ordered moments observed for $T_1 < T < T_2$ in terms of a model with strong orbital fluctuations. This interpretation is supported by the results of inelastic neutron scattering measurements as well as model calculations. Here we point out an additional aspect that has thus far not been noticed. The group-theoretical analysis presented in the Appendix shows that the observed magnetic structure is incompatible with any of the space groups that have been proposed as a description of the crystal structure in the intermediate-temperature phase (that is, the orthorhombic space group *Pbnm* and the monoclinic space groups *P2$_1$/b* and *Pb* discussed in Section III.A), if only terms bilinear in the spin operators are included. For the vast majority of magnetically ordered materials, an analysis based on spin Hamiltonians of this form is adequate, because biquadratic and higher-order terms are negligible. This is also the case for the low-temperature phase of YVO$_3$ and the magnetically ordered states of NdVO$_3$ and TbVO$_3$ (see below). The failure of this analysis for the intermediate-temperature phase of YVO$_3$ indicates that terms beyond the usual bilinear exchange coupling in the spin Hamiltonian are unusually large. In fact, if biquadratic terms are included in the spin Hamiltonian, the crystal and magnetic structures are fully compatible (see Appendix). The presence of such higher-order terms is another fingerprint of the unusual magnetic state realized in YVO$_3$ in the temperature range $T_1 < T < T_2$. A quantitative assessment of the implications of this observation will have to await detailed theoretical work. On a qualitative level, large higher-order terms may well be generic consequences of the near-degeneracy of the $t_{2g}$ levels, which is particularly pronounced in the intermediate-temperature phase of YVO$_3$. For rigid orbitals with a large crystal field splitting, the Heisenberg exchange coupling is simply a scalar multiplying the bilinear term in the spin Hamiltonian. If the crystal field gap is small or absent, and substantial fluctuations are present in the orbital sector, the exchange coupling itself is a quantum variable. This naturally leads to higher-order terms in a free-energy expansion.

As already mentioned above, thermodynamic experiments suggested that the low-temperature structural and magnetic phase transition, which is observed for the yttrium compound at $T_1 = 77$ K, is absent for vanadates with an ionic size larger than that of Dy$^{3+}$.[22] Our neutron diffraction experiments on NdVO$_3$ and TbVO$_3$ fully confirm this conclusion. In fact, the temperature dependence of the intensity of some prominent magnetic reflections of these compounds, shown in Figs. 5 and 6, shows no indications of phase transitions beside the Néel transitions at 135 K and 111 K, respectively, and the spontaneous ordering of the Tb moments at about 11 K, which will be addressed below. The presence of both the 100 and 010 reflections suggests a *C*-type ordering of the V-moments. We found similar intensity ratios for the reflections 100 and 102 for the neodymium and terbium compounds, indicating an order of the vanadium moments within the *ab*-plane. This is also the dominant magnetic mode observed in the intermediate-temperature phase of YVO$_3$. However, for both the neodymium and terbium compounds the reflection 011 remains absent down to 10 K, indicating the absence of the $G_z$ mode observed in YVO$_3$. The unusual canted structure we found in the intermediate-temperature phase of YVO$_3$ therefore appears to be characteristic of compounds that exhibit a phase transition into a *G*-type antiferromagnetic state at low temperatures.

For NdVO$_3$ and TbVO$_3$ the magnetic intensities exhibit a nonmonotonic temperature dependence (Figs. 5 and 6). This suggests the influence of the neodymium and terbium moments, which are polarized in the presence of the exchange field produced by the saturated antiferromagnetic vanadium sublattice. We have therefore first determined the magnitude and orientation of the vanadium moments at high temperatures, where this influence is weak. This was accomplished on the basis of data sets collected with the neutron wavelength $\lambda = 2.4$ Å. In order to obtain the moment values, we determined the overall scale factor from a data set collected at room temperature, where both compounds crystallize in the orthorhombic space group *Pbnm*. The relatively large wavelength only allowed us to collect a small number of Bragg reflections. Therefore the structure refinement of TbVO$_3$ resulted in atomic parameters with significantly larger standard deviations than that of the data set collected at the wavelength $\lambda = 0.89$ Å, but the refined values of both data sets were in a good agreement. The structure refinements of NdVO$_3$ resulted in positional parameters similar to those obtained for the terbium compound, but with the tendency that the tiltings of the VO$_6$-octahedra in the neodymium compound are less pronounced than in TbVO$_3$.

From a data set on NdVO$_3$ collected at 88 K, we thus determined the components $\mu_x = 0.48(2)$ $\mu_B$ and $\mu_y = 0.79(2)$ $\mu_B$ for the antiferromagnetically ordered vanadium sublattice. These values are similar to $\mu_x = 0.49(3)$ $\mu_B$ and $\mu_y = 0.89(2)$ $\mu_B$ obtained for YVO$_3$ at 85 K. For the terbium compound we obtained the magnetic moments of the vanadium ions $\mu_x = 0.83(7)$ $\mu_B$ and $\mu_y = 0.94(7)$ $\mu_B$ from a data set collected at 73 K. Here the value of the *x*-component is somewhat larger than that of the other two compounds. This may reflect the residual influence of the strongly magnetic Tb$^{3+}$-ions and their strong single-ion anisotropy. The total ordered moments of NdVO$_3$ and TbVO$_3$ at 88 K (1.02 $\mu_B$ and 1.25 $\mu_B$, respectively) are far below the free-ion value of 2 $\mu_B$, as already observed in the intermediate-temperature phase of YVO$_3$. This may again indicate substantial orbital fluctuations.

For lower temperatures, the magnetic moments of the Nd and Tb sublattices are polarized in magnetic modes compatible with the magnetic order of the V moments. This was confirmed for NdVO$_3$ from the analysis of our single-crystal neutron data collected at 9.5 K. Here it could be seen that both the V$^{3+}$- and Nd$^{3+}$- sublattices show a pure C-type ordering, where the moments are ordered antiferromagnetically in the ab-plane and ferromagnetically along the c-axis. The magnetic moments of the metal atoms in NdVO$_3$ obtained by a refinement of these data are given in Table III. The exchange coupling between Nd and V spins is antiferromagnetic along the x- and y-directions, resulting in a noncollinear magnetic structure (Fig. 7).

In contrast to the neodymium compound, the magnetic intensity of the reflections 100 and 010 of TbVO$_3$ decreases at lower temperature (Fig. 6). Our data analysis shows that this is because the magnetic moments of the Tb$^{3+}$- and V$^{3+}$-ions are coupled ferromagnetically along the x- and y-directions, whereas those of the Nd$^{3+}$- and V$^{3+}$-ions in NdVO$_3$ are coupled antiferromagnetically. Below 11 K the magnetic intensity of the reflection 010 sharply increases in magnitude. This can be ascribed to the spontaneous magnetic order of the terbium sublattice. From the absence of additional magnetic intensity on the 100 position, we infer a C-type ordering of the terbium moments in the x-direction. A strong increase of the magnetic intensity could also be detected at 200, whereas the change of the intensity for the 020 is negligible. This clearly suggests a ferromagnetic order of the terbium moments along the y-direction. The absence of 001 finally excluded the presence of the modes $A_x$ and/or $A_y$. We also found additional but weaker intensities for the reflections 111, 201 and 021 indicating A modes, as well as for 101 and 011 indicating G-modes. However, these additional reflections are present only because the terbium atoms occupy the general position x, y, ¼. They carry no magnetic intensity when x(Tb) and y(Tb) are equal to zero. Thus we finally could use 30 magnetic reflections for the data analysis. The refinement of the moments along z, assuming the A-mode, resulted in a value close to zero. From refinements of the modes $C_x$ and $F_y$, the magnetic moments of the terbium moments in the ab-plane were calculated to be $\mu_x$ = 6.08(11) $\mu_B$ and $\mu_y$ = 5.03(14) $\mu_B$ with a residual of $R_F$ = 0.09. The total moment $\mu_{exp}$ = 7.89(13) $\mu_B$ is somewhat smaller than the theoretical value of the free Tb$^{3+}$-ion, 9.0 $\mu_B$, presumably because the saturation moment is not reached at 9.5 K. The low-temperature magnetic structure of TbVO$_3$ is shown in Fig. 7. Our results are not in agreement with the magnetic structure given in Ref. 13, where the modes $A_x$ and $G_y$ were reported.

To summarize the rare-earth magnetism in NdVO$_3$ and TbVO$_3$, we note that the magnetically ordered vanadium sublattice induces, via Nd-V and Tb-V exchange, a polarization of the lanthanide moments resulting in a magnetic structure with the same symmetry, so that one finds a pure *C*-type order for both metal sublattices in the *ab*-plane over a wide temperature range below the Néel transitions. A comparable polarization process was observed earlier for other perovskites like NdCrO$_3$, NdFeO$_3$ and NdNiO$_3$,[34-36] where it proved to be amenable to a mean-field description[35]. We point out two additional aspects. First, the signs of the V-Nd and V-Tb exchange interactions are opposite. The same trend was found earlier for the intermetallic compounds Nd$_2$Co$_{12}$P$_7$ and Ho$_2$Co$_{12}$P$_7$, where the induced moments of the (light) neodymium and (heavy) holmium ions are aligned antiparallel and parallel to well ordered cobalt moments, respectively.[37] Conversely, the induced alignment between the Fe and Nd spins in Nd$_2$Fe$_{14}$B is parallel, and that of the Fe and Ho spins in Nd$_2$Fe$_{14}$B is antiparallel.[38] This effect can be attributed to the opposite sign of the spin-orbit coupling parameter for lanthanide ions with less-than-half-filled and more-than-half-filled *f*-electron shells. A second notable aspect is the large polarization of the Nd moments, which significantly exceeds that of the much larger Tb moments at comparable temperatures (Figs. 5 and 6). This indicates that the V-Nd coupling is relatively large.

At 11 K, Tb-Tb interactions drive a spontaneous re-ordering of the terbium moments from a $C_y$ to an $F_y$ mode. Analogous ordering phenomena have been observed in the intermetallic compounds Pr$_2$Co$_{12}$P$_7$[37] and PrCo$_2$P$_2$[39] as well as in layered cuprates such as Nd$_2$CuO$_4$[40,41] Sm$_2$CuO$_4$[42] and Gd$_2$CuO$_4$[43]. In NdVO$_3$, spontaneous magnetic order on the Nd sublattice was not observed above 9.5 K, indicating that the Nd-Nd interactions are not unusually large. The neodymium sublattice of other members of the Nd*B*O$_3$ series exhibit magnetic transitions at about 1 K. This is the case for the compounds NdScO$_3$[44] or NdGaO$_3$[45] containing nonmagnetic *B* atoms, as well as for NdFeO$_3$[46] or NdNiO$_3$[36], where the *B* atoms exhibit magnetic order at 690 K and 200 K, respectively. The cooperative order of the Nd sublattice can also be inhibited through strong effective *R-B* interactions, as shown for NdCrO$_3$.[34] Further work at lower temperature is required to show whether such mechanisms are at work in NdVO$_3$ as well.

## IV. CONCLUSIONS

Motivated by the unusual thermodynamic behavior and spin dynamics of YVO$_3$, we have carried out a comprehensive investigation of the crystal and magnetic structures of this material using single-crystal neutron diffraction as well as neutron and high-resolution x-ray powder diffraction. This material exhibits three structural phases: For both $T > 200$K and $T < 77$ K, the crystal structure is described by the orthorhombic space group *Pbnm*. However, while in the high-temperature phase the lattice distortions away from the ideal cubic perovskite structure are determined entirely by ionic-size effects, a strong cooperative Jahn-Teller effect is active in the low-temperature phase. The diffraction pattern in the intermediate-temperature phase is described by a monoclinic space group, but the monoclinic distortion is extremely small and could not be resolved despite a dedicated effort with high-resolution synchrotron x-ray diffraction. As a consequence, the vanadium positions in this phase could not be refined with sufficient accuracy to ascertain whether the V-O-V bond length alternates along the *c*-axis direction, as hypothesized based on an analysis of neutron spectroscopy data [15] in terms of an "orbital Peierls state" as well as infrared spectroscopy data [32]. The correlations between vanadium $t_{2g}$ orbitals that are presumably responsible for the unusual magnetic dynamics in this phase thus appear to be only very weakly coupled to the crystal lattice. Orbital correlations that are only weakly manifested in the crystal lattice have also been observed in other $t_{2g}$ systems such as Ca$_2$RuO$_4$. [47]

While the *G*-type magnetic structure of YVO$_3$ in the low-temperature phase is collinear and fully compatible with the orthorhombic crystal structure, the magnetic structure in the intermediate-temperature phase is highly noncollinear. This may be a manifestation of the unusual orbital correlations in this phase. The group theoretical analysis presented in the Appendix has revealed that this magnetic structure is incompatible with the crystal structure, unless the spin Hamiltonian is expanded beyond the usual bilinear terms. The presence of substantial higher-order spin-spin interactions has not been predicted. Theoretical work is required to assess whether they could be another consequence of the unusual orbital correlations in the intermediate-temperature phase of YVO$_3$.

The properties of YVO$_3$ were contrasted to those of TbVO$_3$ and NdVO$_3$, whose room temperature lattice structure is closely similar. These materials differ from YVO$_3$ in two respects. First, whereas the Y$^{3+}$ ion is nonmagnetic, the Tb$^{3+}$ and Nd$^{3+}$ ions carry substantial magnetic moments. The exchange interactions between rare-earth and vanadium moments are weaker than the vanadium-vanadium interactions and lead to a polarization of the rare-earth moments. At low temperatures, direct interactions between the rare-earth moments lead to a spontaneous ordering of the Tb-sublattice in TbVO$_3$. A similar transition is expected in NdVO$_3$, but was not observed due to the limited temperature range covered by our experiments. While the magnetic structures we have found are at variance with some previous reports, they are in general accord with other materials containing both *d*- and *f*-electron elements.

The second major difference between YVO$_3$ on the one hand, and TbVO$_3$ and NdVO$_3$ on the other hand, is the absence of the low-temperature orthorhombic phase with *G*-type magnetic structure in the latter two materials. In both of these compounds, the vanadium moments are arranged in a *C*-type pattern at low temperature, similar to the magnetic structure of the intermediate-temperature phase in YVO$_3$. The large spin canting observed in YVO$_3$ at these temperatures is, however, missing in TbVO$_3$ and NdVO$_3$, and the crystal and magnetic structures are fully compatible without taking higher-order terms in the spin Hamiltonian into account. It remains to be seen whether this observation is compatible with extant models of the electronic properties of the pseudocubic vanadates. These and other results reported here thus broaden the experimental basis for a quantitative understanding of the unusual orbital and spin correlations in the vanadates.

# APPENDIX: GROUP THEORETICAL DESCRIPTION OF POSSIBLE MAGNETIC STRUCTURES

In the space group *Pbnm* the vanadium atoms are on the Wyckoff position 4*b*: (1) ½, 0, 0; (2) ½, 0, ½; (3) 0, ½, 0; (4) 0, ½, ½. As the operation ***I*** does not change anything we end up only with four irreducible representations $\Gamma_j$ ($j$ = 1, 2, 3, 4) as shown in Table IV. From our experiment we found for the vanadium moments in YVO$_3$ the modes $C_x$, $C_y$ and $G_z$. It is interesting to see, that only the magnetic ordering in the *bc*-plane is compatible with *Pbnm* as deduced from Bertaut's representation analysis [48] resulting in the basis function [$F_x$, $C_y$, $G_z$]. The magnetic structure of the low-temperature phase is thus fully compatible with this analysis. In the intermediate-temperature phase, however, we found an antiferromagnetic coupling (*C*-type) along the *a*-axis, instead of a ferromagnetic one. In order to check whether the incompatibility between the crystal and magnetic structure can be explained by the fact that the intermediate phase of YVO$_3$ crystallizes in the monoclinic space group *P*2$_1$/*b* (*P* 2$_1$/*b* 1 1), as discussed in the text, we carried out a similar analysis for this space group. Here, *a* is the unique axis, and there are 4 symmetry operators $E$: $(x, y, z)$; $2_1x$: $(x + ½, -y + ½, -z)$; $I$: $(-x, -y, -z)$; $b$: $(-x + ½, y + ½, z)$. All the symmetry operators commute, which means that there are four one-dimensional irreducible representations $\Gamma_j$ ($j = 1, 2, 3, 4$). In $P2_1/b$ the vanadium atoms are in the different Wyckoff positions $2c$: (½, 0, 0 and 0, ½, 0) and $2d$ (½, 0, ½ and 0, ½, ½). A symmetry analysis finally results in two irreducible representations $\Gamma_1$ and $\Gamma_3$ with the basis functions $[F_x, A_y, A_z]$ and $[A_x, F_y, F_z]$, while the others belonging to $\Gamma_2$ and $\Gamma_4$ are null (Table IV). The vector sum $F = S_1 + S_2$ for each site in $P2_1/b$ corresponds to $S_1 + S_3$ and $S_2 + S_4$ in $Pbnm$, respectively. An antiferromagnetic coupling between the atoms in $2c$ and $2d$ in $P2_1/b$ (or the coupling of the atoms along the $c$-axis) corresponds to the mode $A(+ - + -)$ in $Pbnm$. Alternatively, with the vector sum $A = S_1 - S_2$ for each site in $P2_1/b$ corresponds to $S_1 - S_3$ and $S_2 - S_4$ in $Pbnm$, respectively. In the case of a ferromagnetic coupling between the atoms in $2c$ and $2d$ we finally obtain the observed mode $C(+ + - -)$ and in the case of an antiferromagnetic coupling the observed mode $G(+ - - +)$. It can be seen that no combination can satisfy the observed modes $C_x$, $C_y$ and $G_z$. Further we carried out the symmetry analysis for $Pb$ ($P b 1 1$), as given by Tsvetkov et al. [32] In this space group there are only two symmetry elements: $E$: $(x, y, z)$ and $b$: $(-x, y + ½, z)$ giving two one-dimensional irreducible representations. By the same procedure we get the same basis vectors as in $P2_1/b$. Consequently, the modes $C_x$, $C_y$ and $G_z$ are incompatible if one takes only one irreducible representation. However, this configuration can be explained by taking a combination of two irreducible representations. This would mean that terms of higher order than the usual bilinear spin-spin interactions determine the magnetic ground state [48]

The magnetic structure of the terbium sublattice with the modes $C_x$ and $F_y$ can be deduced from Bertaut's representation analysis [48] using the orthorhombic space group $Pbnm$ and the propagation vector $k = 0$ as presented in Table V. In this space group, the irreducible representation of the group $G_k$ are those of the point group $G_0 = mmm$ ($D_{2h}$). $G_0$ contains 8 symmetry elements, labelled as 1: $E$ (identity), 2: $2_1x$; 3: $2_1y$; 4: $2_1z$; 5: $I$(inversion); 6: $n$; 7: $m$; 8: $b$. Since all the symmetry operators commute, all the irreducible representations are one-dimensional and labelled $\Gamma_1 \ldots \Gamma_8$. Table V summarizes the transformation properties of the modes $F(+ + + +)$, $C(+ + - -)$, $G(+ - - +)$ and $A(+ - + -)$ of the Tb atoms located in the Wyckoff position $4c$: (1) $x, y, ¼$; (2) $-x, -y, ¾$; (3) $½ - x, ½ + y, ¼$; (4) $½ + x, ½ - y, ¾$. The type of ordering of the terbium moments is surprising, because a transition from the orthorhombic into the monoclinic structure already sets in for TbVO$_3$ at about 190 K and it remains stable down to 10 K. As already discussed above, the monoclinic distortions arise from the *Jahn-Teller* effect resulting predominantly in a change to point symmetry of the vanadium ions according to the monoclinic space group *P*2$_1$/*b*. The influence of the superposed distortions on the terbium sites seems to be negligible and the magnetic ordering is still compatible with the symmetry of *Pbnm*.

## ACKNOWLEDGMENTS

We acknowledge valuable discussions with G. Khaliullin and the support of the Deutsche Forschungsgemeinschaft under grant UL 164/4.

TABLE I. Results of the structure refinements of YVO$_3$ and TbVO$_3$. For the single-crystal and powder data collected at 5 K, 85 K, 230 K and 295 K the crystal structures were refined in the orthorhombic space group *Pbnm*. Additionally the crystal structure of YVO$_3$ at 85 K was refined in the monoclinic space group *P2$_1$/b*. Here the *z* value of the yttrium atom was not allowed to vary during the refinements. The thermal parameters $U_{ij}$ (given in 100 Å$^2$) are in the form $\exp[-2\pi^2(U_{11} h^2 a^{*2} + \ldots 2U_{13} h l a^* c^*)]$. For symmetry reasons the values $U_{13}$ and $U_{23}$ of the Y-, Tb- and O1-atoms are equal to zero for the space group *Pbnm*. In the case of the powder diffraction study of YVO$_3$ only the isotropic thermal parameters were refined.

| | YVO$_3$ at 5 K † | YVO$_3$ at 85 K † | YVO$_3$ at 85 K † | YVO$_3$ at 230 K † | YVO$_3$ at 295 K ‡ | TbVO$_3$ at 295 K ‡ |
|---|---|---|---|---|---|---|
| Space group | *Pbnm* | *Pbnm* | *P2$_1$/b* | *Pbnm* | *Pbnm* | *Pbnm* |
| *a* [Å] | 5.28551(6) | 5.28547(7) | 5.27650(5) | 5.27953(8) | 5.27722(3)* | 5.3206(10) |
| *b* [Å] | 5.59264(5) | 5.62399(7) | 5.62401(5) | 5.61072(8) | 5.60453(3)* | 5.6068(11) |
| *c* [Å] | 7.55615(7) | 7.53979(11) | 7.53980(7) | 7.57214(13) | 7.57294(4)* | 7.6035(15) |
| *α* [°] | 90 | 90 | 89.980 * | 90 | 90 | 90 |
| *x* (Y/Tb) | 0.9783(2) | 0.9797(2) | 0.9798(2) | 0.9797(3) | 0.98077(9) | 0.98258(12) |
| *y* (Y/Tb) | 0.0715(2) | 0.0712(2) | 0.0713(2) | 0.0713(3) | 0.06905(10) | 0.06497(12) |
| *z* (Y/Tb) | 0.25 | 0.25 | 0.2500 | 0.25 | 0.25 | 0.25 |
| *x* (O1) | 0.1126(2) | 0.1105(2) | 0.1106(2) | 0.1117(3) | 0.11056(12) | 0.10443(16) |
| *y* (O1) | 0.4622(3) | 0.4623(2) | 0.4620(3) | 0.4618(3) | 0.46017(9) | 0.46439(16) |
| *z* (O1) | 0.25 | 0.25 | 0.2507(19) | 0.25 | 0.25 | 0.25 |
| *x* (O2) | 0.6872(2) | 0.6907(2) | 0.6938(13) | 0.6908(2) | 0.69123(9) | 0.69399(11) |
| *y* (O2) | 0.2996(3) | 0.3040(2) | 0.3110(7) | 0.3040(2) | 0.30394(10) | 0.30175(11) |
| *z* (O2) | 0.05663(10) | 0.05610(11) | 0.0545(11) | 0.05563(12) | 0.05630(6) | 0.05352(8) |
| *x* (O3) | | | 0.3124(14) | | | |
| *y* (O3) | | | 0.7026(7) | | | |
| *z* (O3) | | | 0.5579(11) | | | |
| $U_{11}$ (Y/Tb) | 0.28(2) | 0.36(2) | 0.23(3) | 0.53(3) | 0.46(2) | 0.61(3) |
| $U_{22}$ (Y/Tb) | | | | | 0.41(2) | 0.38(2) |
| $U_{33}$ (Y/Tb) | | | | | 0.52(2) | 0.39(2) |
| $U_{12}$ (Y/Tb) | | | | | −0.06(1) | −0.05(2) |
| $U_{is}$ (V) | 0.3 | 0.3 | 0.3 | 0.3 | 0.3 | 0.3 |
| $U_{11}$ (O1) | 0.28(3) | 0.36(3) | 0.49(3) | 0.47(3) | 0.54(2) | 0.81(3) |
| $U_{22}$ (O1) | | | | | 0.62(3) | 0.66(3) |
| $U_{33}$ (O1) | | | | | 0.39(2) | 0.37(3) |
| $U_{12}$ (O1) | | | | | −0.11(2) | −0.08(3) |
| $U_{11}$ (O2, O3) | 0.33(2) | 0.43(2) | 0.49(2) | 0.52(2) | 0.52(2) | 0.73(2) |
| $U_{22}$ (O2) | | | | | 0.55(2) | 0.49(2) |
| $U_{33}$ (O2) | | | | | 0.64(2) | 0.67(2) |
| $U_{12}$ (O2) | | | | | −0.09(1) | −0.10(2) |
| $U_{13}$ (O2) | | | | | 0.09(1) | 0.10(2) |
| $U_{23}$ (O2) | | | | | −0.11(1) | −0.12(2) |

† from neutron powder data, ‡ from single-crystal neutron data, * from synchrotron powder data

TABLE II. Lattice constants, cell volumes and angles (in Å, in Å$^3$ and °) as well as the interatomic distances (in Å) in the $T$O$_6$-octahedron with $T$ = V, Fe. For the data sets YVO$_3$ of collected at 85 K the structure refinements were carried out in the space groups *Pbnm* and *P2$_1$/b*. The values of YVO$_3$ and TbVO$_3$ as determined in this work are compared with those of PrFeO$_3$ and LuFeO$_3$ given in Ref. 30.

|  | YVO$_3$ | | | | | TbVO$_3$ | PrFeO$_3$ | LuFeO$_3$ |
|---|---|---|---|---|---|---|---|---|
| $T$ | 5 K [a] | 85 K [a] | 85 K [a] | 230 K [a] | 295 K [b] | 295 K [c] | RT [d] | RT [d] |
| Space group | *Pbnm* | *Pbnm* | *P2$_1$/b* | *Pbnm* | *Pbnm* | *Pbnm* | *Pbnm* | *Pbnm* |
| $d_{T1-O11}$ | 1.9919(4) | 1.9845(5) | 1.990(13) | 1.9943(5) | 1.9936(2) | 1.9905(4) | 2.001(1) | 2.008(2) |
| $d_{T1-O21}$ | 2.0427(12) | 2.0143(11) | 1.977(6) | 2.0129(12) | 2.0110(5) | 2.0130(7) | 2.009(3) | 1.997(4) |
| $d_{T1-O22}$ | 1.9925(13) | 2.0283(11) | 2.067(5) | 2.0251(12) | 2.0253(5) | 2.0232(7) | 2.015(3) | 2.024(4) |
| $d_{T2-O11}$ | | | 1.980(13) | | | | | |
| $d_{T2-O31}$ | | | 1.992(5) | | | | | |
| $d_{T2-O32}$ | | | 2.050(6) | | | | | |
| $a$ | 5.28551(6) | 5.27647(7) | 5.27650(5) | 5.27953(8) | 5.27722(1) | 5.3206(10) | 5.482(1) | 5.213(3) |
| $b$ | 5.59264(5) | 5.62399(7) | 5.62401(5) | 5.61072(8) | 5.60454(1) | 5.6068(11) | 5.578(1) | 5.547(3) |
| $c$ | 7.55615(7) | 7.53979(11) | 7.53980(7) | 7.57214(13) | 7.57294(2) | 7.6035(15) | 7.786(1) | 7.565(3) |
| $\alpha$ | 90 | 90 | 89.980 [e] | 90 | 90 | 90 | 90 | 90 |
| $V$ | 223.359(7) | 223.742(9) | 223.744(7) | 224.302(10) | 223.980(1) | 226.82(13) | 238.08(12) | 218.8(3) |

[a] from neutron powder data

[b] $d_{V-O}$ values from single-crystal neutron data; lattice constants from high-resolution synchrotron powder data

[c] from single-crystal neutron data

[d] from single-crystal x-ray data

TABLE III. Magnetic moments of the V- and Nd-atoms in NdVO$_3$ at 9.5 K as obtained from the refinements.

| Nd | $\mu_x$ [$\mu_B$] | $\mu_y$ [$\mu_B$] | V | $\mu_x$ [$\mu_B$] | $\mu_y$ [$\mu_B$] |
|---|---|---|---|---|---|
| $x, y, ¼$ | +0.23(3) | −0.47(4) | ½, 0, 0 | +0.56(16) | +0.90(4) |
| $−x, −y, ¾$ | +0.23(3) | −0.47(4) | ½, 0, ½ | +0.56(16) | +0.90(4) |
| $½ − x, ½ + y, ¼$ | −0.23(3) | +0.47(4) | 0, ½, 0 | −0.56(16) | −0.90(4) |
| $½ + x, ½ − y, ¾$ | −0.23(3) | +0.47(4) | 0, ½, ½ | −0.56(16) | −0.90(4) |

TABLE IV. Representations of the base vectors of $\{T\}$ in $RTO_3$ for the space groups $Pbnm$ and $P2_1/b$, using the propagation vector $\mathbf{k} = 0$. In the monoclinic $P2_1/b$ two different atoms V1 and V2 are located in the Wyckoff positions $2c$ (½, 0, 0 and 0, ½, 0) and $2d$ (½, 0, ½ and 0, ½, ½), respectively. The modes $A$ and $F$ mean a spin succession of $+ -$ and $+ +$. In the orthorhombic $Pbnm$ the vanadium atoms are located on the same positions, but on the Wyckoff position $4b$ (½, 0, 0; ½, 0, ½; 0, ½, 0; 0, ½, ½). The different modes are $F(+++ +)$, $C(++--)$, $G(+--+)$ and $A(+-+-)$.

| Pbnm | x | y | z | $P2_1/b$ | x | y | z |
|---|---|---|---|---|---|---|---|
| $\Gamma_1$ | $A_x$ | $G_y$ | $C_z$ | $\Gamma_1$ | $F_x$ | $A_y$ | $A_z$ |
| $\Gamma_2$ | $F_x$ | $C_y$ | $G_z$ | $\Gamma_2$ | - | - | - |
| $\Gamma_3$ | $C_x$ | $F_y$ | $A_z$ | $\Gamma_3$ | $A_x$ | $F_y$ | $F_z$ |
| $\Gamma_4$ | $G_x$ | $A_y$ | $F_z$ | $\Gamma_4$ | - | - | - |

TABLE V. Representations of the base vectors of $\{R\}$ in $RTO_3$ with the space group $Pbnm$ and the propagation vector $\mathbf{k} = 0$. The $R$ atoms are located in the Wyckoff position $4c$ ($x$, $y$, ¼; $-x$, $-y$, ¾; ½ $- x$, ½ $+ y$, ¼; ½ $+ x$, ½ $- y$, ¾). The modes are $F(+++ +)$, $C(++--)$, $G(+--+)$ and $A(+-+-)$.

| Pbnm | x | y | z |
|---|---|---|---|
| $\Gamma_1$ | – | – | $C_z$ |
| $\Gamma_2$ | $F_x$ | $C_y$ | – |
| $\Gamma_3$ | $C_x$ | $F_y$ | – |
| $\Gamma_4$ | – | – | $F_z$ |
| $\Gamma_5$ | $G_x$ | $A_y$ | – |
| $\Gamma_6$ | – | – | $A_z$ |
| $\Gamma_7$ | – | – | $G_z$ |
| $\Gamma_8$ | $A_x$ | $G_y$ | – |

**Figure captions**

FIG. 1. Projection of the orthorhombic structure of YVO$_3$ along the $c$-axis. The large circles are yttrium, the medium circles oxygen and the small filled circles vanadium. The yttrium atoms are at $z$ = ¼ and $z$ = ¾. In the drawing the network of distorted corner-shared VO$_6$-octahedra in the $ab$-plane is shown. The octahedra are rotated mostly around the $b$- and $c$-axes and slightly around the $a$-axis.

FIG. 2. Lattice parameters of vanadates $R$VO$_3$ with GdFeO$_3$-type structure as obtained from x-ray diffraction at room temperature. [5] The values of the yttrium compound are placed between the dysprosium and holmium compounds with respect to their ionic radii.

FIG. 3. Temperature dependence of some prominent reflections of YVO$_3$. Below 200 K a structural phase transition occurs from the orthorhombic (*Pbnm*) into the monoclinic structure (*P2$_1$/b*). The weak change of the intensity of the reflection 101 at 116 K indicates the *G*-type ordering along the $c$-axis. Below the second structural phase transition at about 75 K returning to the orthorhombic phase (*Pbnm*), a pure $G_z$-type ordering is observed.

FIG. 4. Rietveld refinements of the neutron diffraction data of YVO$_3$ collected at 5 K. The positions of the nuclear (N) and magnetic reflections (M) as well as the difference pattern (D) are shown. Due to the *G*-type ordering of the vanadium moments, magnetic intensity was found on the positions of the forbidden and allowed reflections 011 and 101, respectively.

FIG. 5. Temperature dependence of the magnetic reflections 010, 100, 012 and 102 of NdVO$_3$. Below about 135 K the magnetic ordering of the vanadium moments begins. The increase of the intensities of 010, 100 and 012, as well the decrease of that of the 102, indicates the additional induced ordering of the neodymium moments.

FIG. 6. Temperature dependence of the magnetic reflections 010 and 100 of TbVO$_3$. In the upper diagrams a logarithmic scale has been used. At 9.5 K the intensity of the reflection 010 strongly increases due to the spontaneous magnetic order of the terbium moments ($C_x$ type). Between 10 K and 40 K the terbium moments are slightly induced by the ordered vanadium moments. The magnetic order of the vanadium moments disappear at about 110 K.

FIG. 7. Magnetic structure of NdVO$_3$ and TbVO$_3$ at 9.5 K. The magnetic moments of the metal atoms are aligned in the *ab*-plane. The large circles are neodymium or terbium, the medium circles oxygen and the small filled circles vanadium. The lanthanide atoms are at $z = ¼$ and $z = ¾$. In the drawing the network of distorted corner-shared VO$_6$-octahedra in the *ab*-plane is shown.

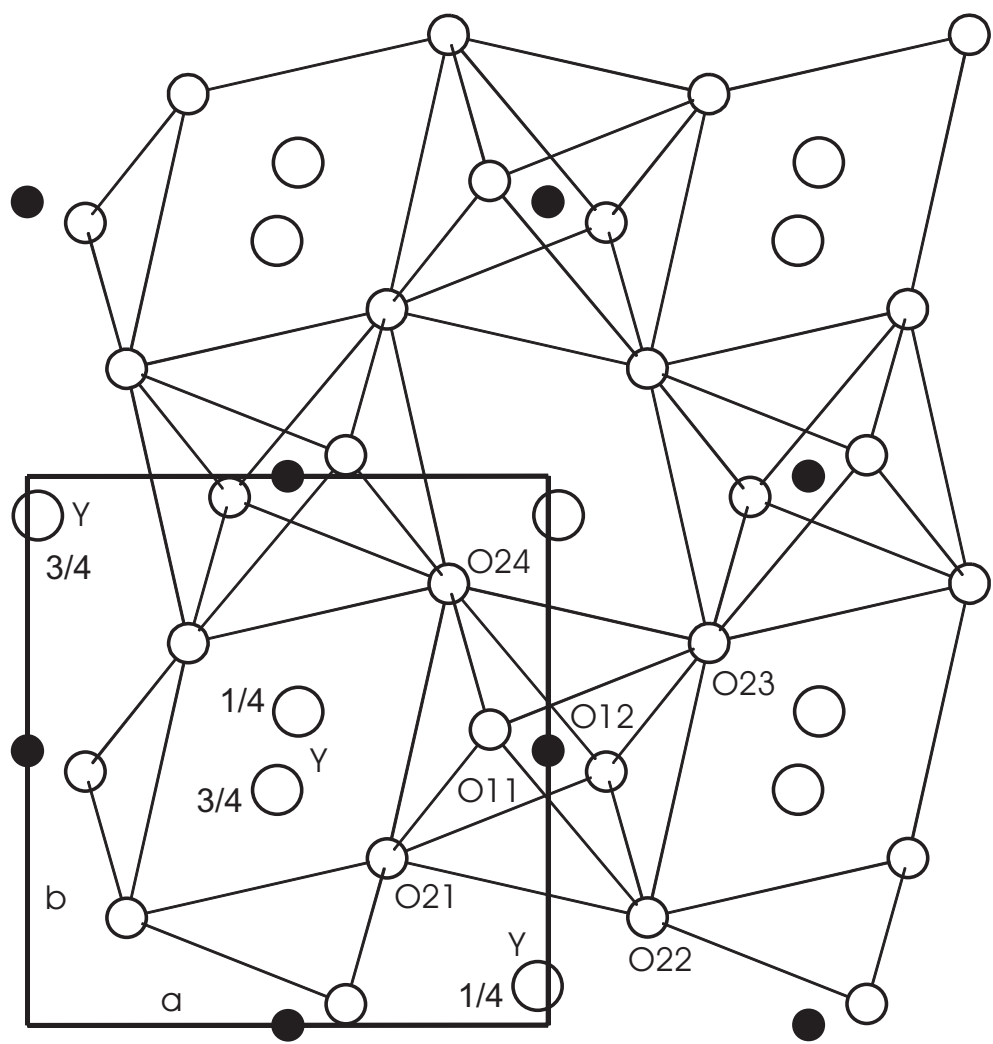

FIG. 1. Projection of the orthorhombic structure of YVO$_3$ along the *c*-axis. The large circles are yttrium, the medium circles oxygen and the small filled circles vanadium. The yttrium atoms are at $z = ¼$ and $z = ¾$. In the drawing the network of distorted corner-shared VO$_6$-octahedra in the *ab*-plane is shown. The octahedra are rotated mostly around the *b*- and *c*-axes and slightly around the *a*-axis.

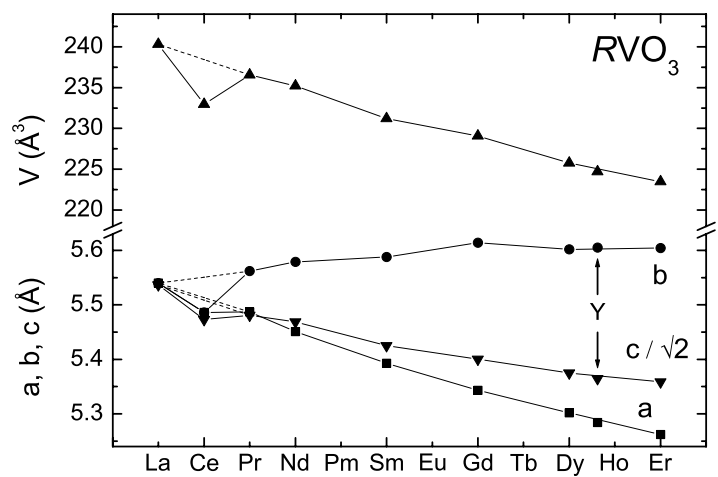

FIG. 2. Lattice parameters of vanadates $R$VO$_3$ with GdFeO$_3$-type structure as obtained from x-ray diffraction at room temperature.[5] The values of the yttrium compound are placed between the dysprosium and holmium compounds with respect to their ionic radii.

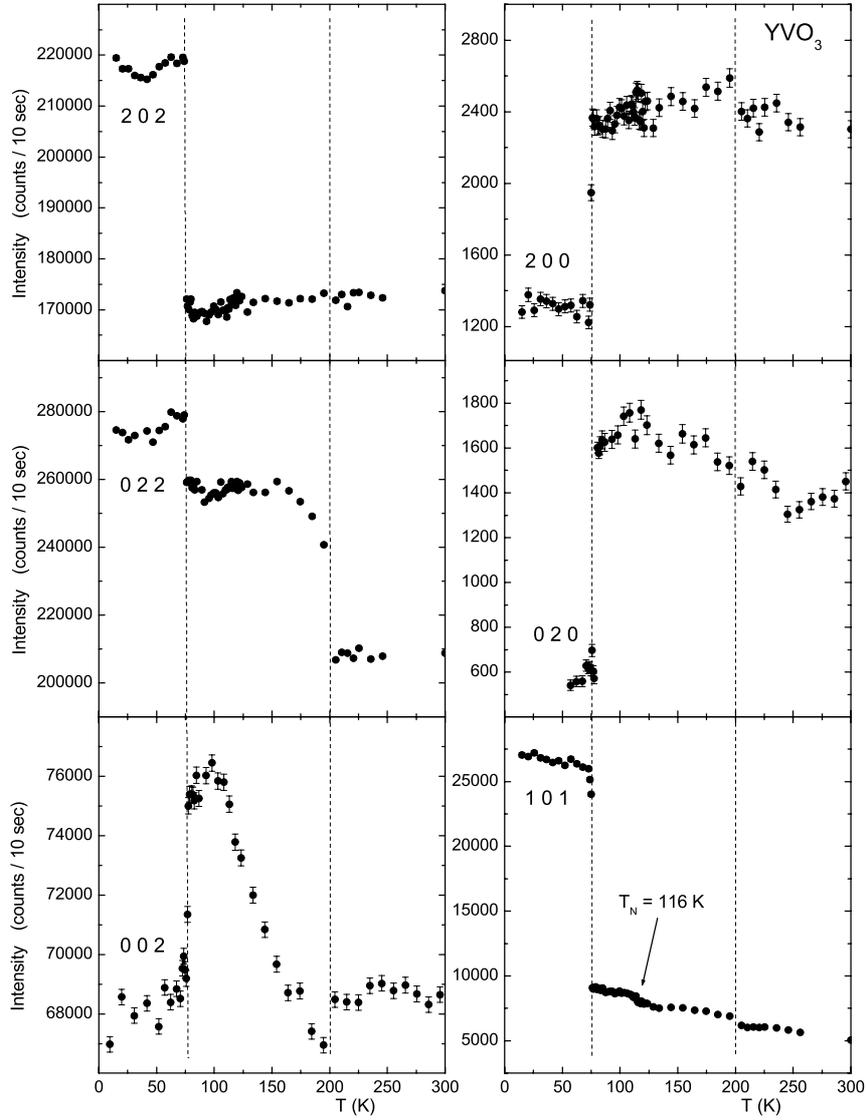

FIG. 3. Temperature dependence of some prominent reflections of YVO$_3$. Below 200 K a structural phase transition occurs from the orthorhombic (*Pbnm*) into the monoclinic structure (*P2$_1$/b*). The weak change of the intensity of the reflection 101 at 116 K indicates the *G*-type ordering along the *c*-axis. Below the second structural phase transition at about 75 K returning to the orthorhombic phase (*Pbnm*) a pure $G_z$-type ordering is observed.

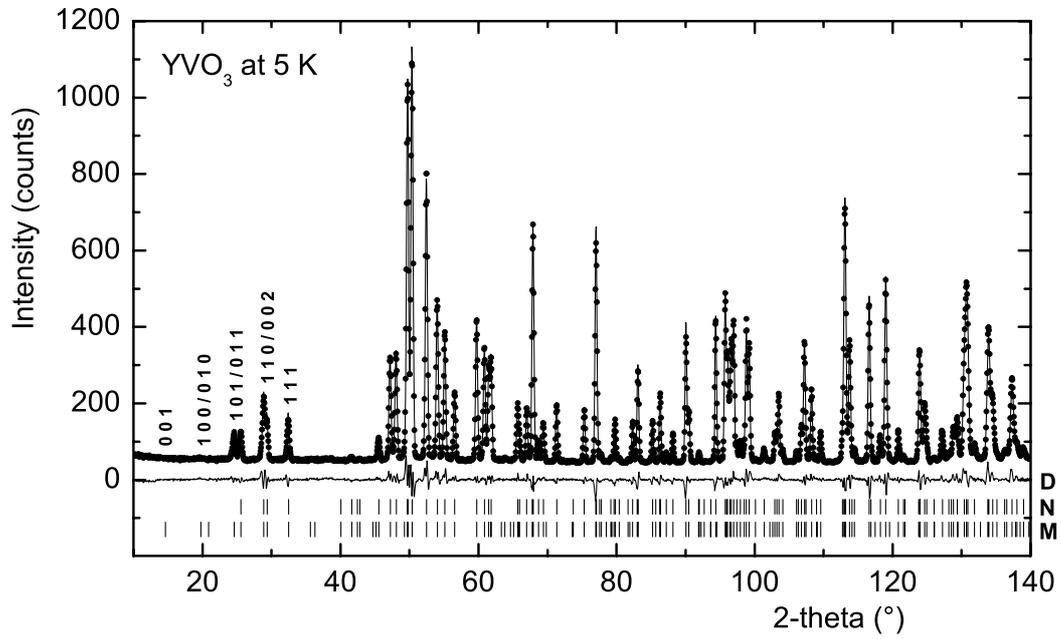

FIG. 4. Rietveld refinements of the neutron diffraction data of YVO$_3$ collected at 5 K. The positions of the nuclear (N) and magnetic reflections (M) as well as the difference pattern (D) are shown. Due to the *G*-type ordering of the vanadium moments, magnetic intensity was found on the positions of the forbidden and allowed reflections 011 and 101, respectively.

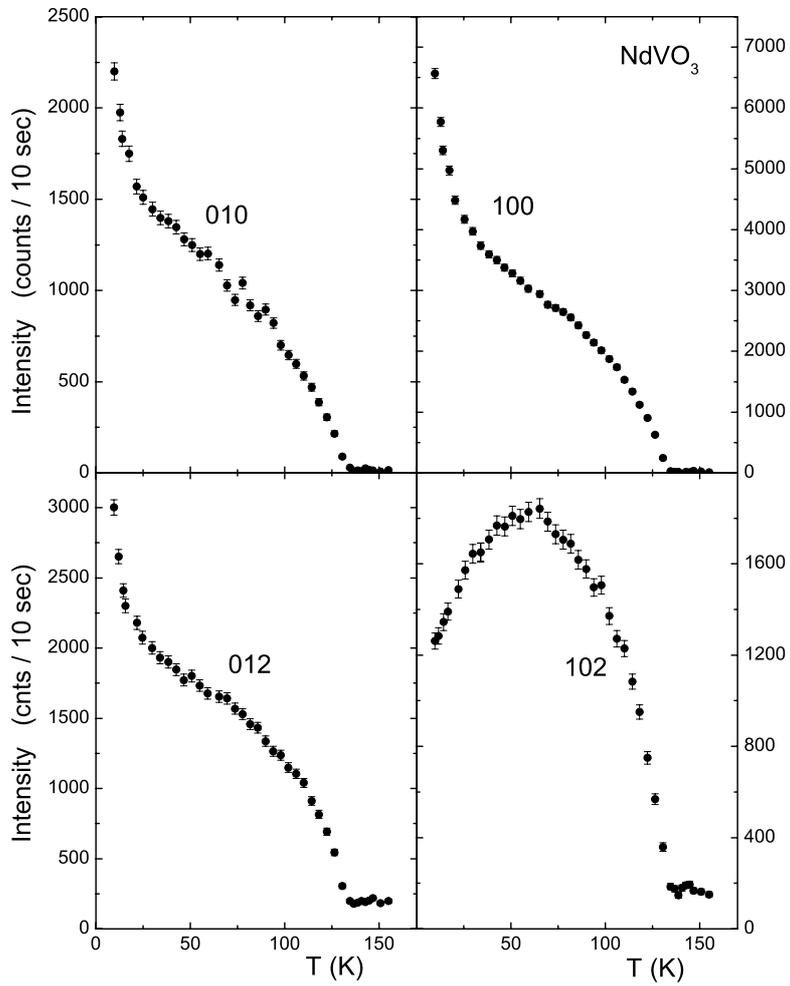

Fig. 5. Temperature dependence of the magnetic reflections 010, 100, 012 and 102 of NdVO$_3$. Below about 135 K the magnetic ordering of the vanadium moments begins. The increase of the intensities of 010, 100 and 012, as well the decrease of that of the 102, indicates the additional induced ordering of the neodymium moments.

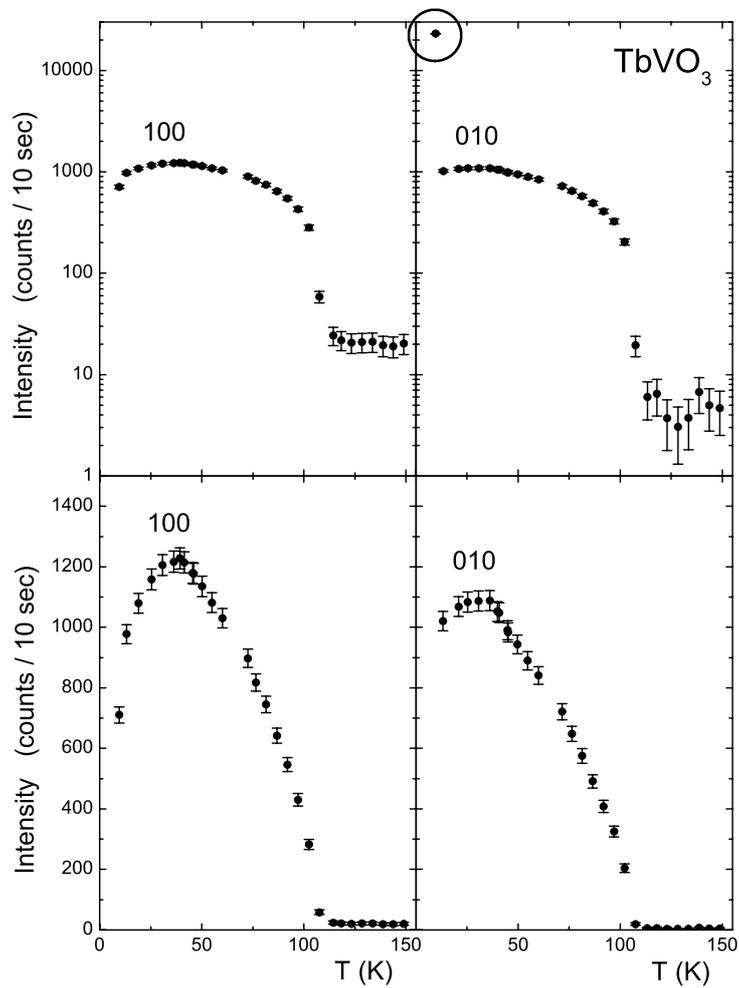

Fig. 6. Temperature dependence of the magnetic reflections 010 and 100 of TbVO$_3$. In the upper diagrams a logarithmic scale has been used. At 9.5 K the intensity of the reflection 010 strongly increases due to the spontaneous magnetic order of the terbium moments ($C_x$ type). Between 10 K and 40 K the terbium moments are slightly induced by the ordered vanadium moments. The magnetic order of the vanadium moments disappear at about 110 K.

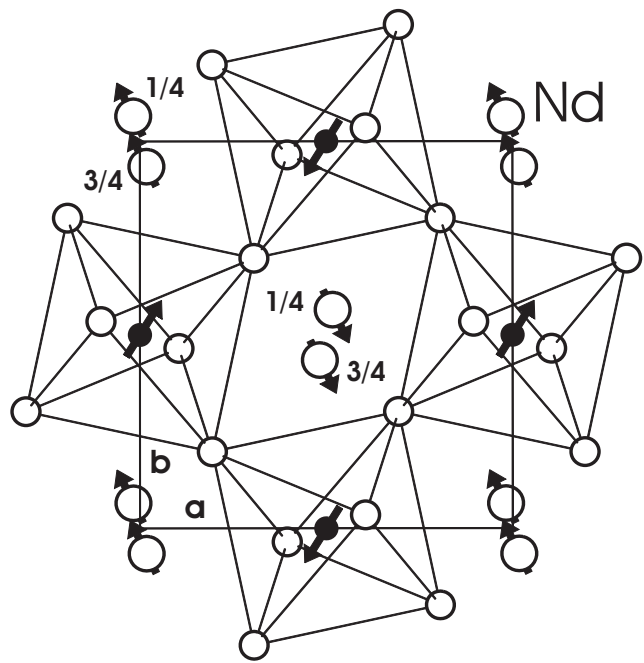

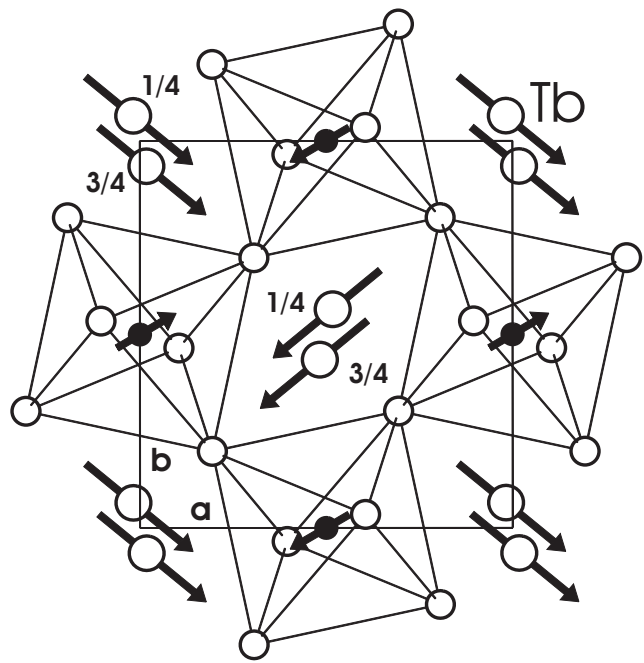

Fig. 7. Magnetic structure of NdVO$_3$ and TbVO$_3$ at 9.5 K. The magnetic moments of the metal atoms are aligned in the *ab*-plane. The large circles are neodymium or terbium, the medium circles oxygen and the small filled circles vanadium. The lanthanide atoms are at $z =$ ¼ and $z =$ ¾. In the drawing the network of distorted corner-shared VO$_6$-octahedra in the *ab*-plane is shown.